\documentclass[%
 reprint,
 amsmath,amssymb,
 aps,
]{revtex4-1}

\usepackage{graphicx}
\usepackage{dcolumn}
\usepackage{bm}
\usepackage{color}
\usepackage{float}

\begin{document}


\title{Cooperation in a generalized age-structured spatial game}

\author{Paulo Victor S Souza}
\email{paulo.victor@ifrj.edu.br}
 \affiliation{Instituto Federal de Educa\c c\~ao, Ci\^encia e Tecnologia do Estado do Rio de Janeiro, Campus Volta Redonda, 27215-350 Volta Redonda, Rio de Janeiro, Brazil}



\author{Rafael Silva}%
\email{raf359@hotmail.com}
\affiliation{
	Departamento de Ci\^encias Exatas e Licenciaturas, Universidade Federal Fluminense, 27213-145, Volta Redonda, Rio de Janeiro, Brazil
}%

\author{Chris T. Bauch}
 \email{cbauch@uwaterloo.ca}
\affiliation{Department of Applied Mathematics, University of Waterloo, 200 University Avenue West, Waterloo, Ontario, Canada N2L 3G1}%
\author{Daniel Girardi}%
\email{d.girardi@ufsc.br}
\affiliation{%
	Departamento de Ci\^encias Exatas e Licenciaturas, Campus Blumenau, Universidade Federal de Santa Catarina, 89036-256, Blumenau, Santa Catarina, Brazil
}%
\date{\today}

\begin{abstract}

The emergence and prevalence of cooperative behavior within a group of selfish individuals remains a puzzle for \text{evolutionary game theory} precisely because it conflicts directly with the central idea of natural selection. Accordingly, in recent years, the search for an understanding of how cooperation can be stimulated, even when it conflicts with individual interest, has intensified. We investigate the emergence of cooperation in an age-structured evolutionary spatial game. In it, players age with time and the payoff that they receive after each round \text{depends on} their age. \text{We find that t}he outcome of the game is strongly influenced by the type of distribution used to modify the payoffs according to the age of each player. The results show that, under certain circumstances, cooperators may not only survive but dominate the population.  
\end{abstract}

\maketitle


\section{INTRODUCTION}

Cooperative behavior is an ubiquitous and widespread phenomenon \text{that} ranges from biological systems to human societies. The reasons behind the emergence, enhancement and maintenance of cooperation remain one of the most important and interesting puzzles in social dilemmas, because there usually is a conflict of interest between what is good for the individual and what is good for the population \cite{nowak2006five,perc2017statistical}. The cooperative behavior manifested between selfish and unrelated individuals are studied \text{in} the field of \text{evolutionary game theory}, established by Smith and Price in 1973 \cite{smith1973lhe}, that provides an appropriate theoretical framework to approach this still open problem. A paradigmatic model frequently employed to investigate the problem of cooperation in evolutionary game theory is the prisoner's dilemma game which was introduced by \text{Merrill Flood and Melvin Dresher in 1950 \cite{rapoport1965prisoner}.} In the basic version of the prisoner's dilemma game,  two players interact and they need to \text{choose between the two strategies of} cooperate or defect. If both defect (cooperate) they obtain the punishment $P$ (the reward $R$). However, when a cooperator is confronted with a defector, the cooperator receives the sucker's payoff $S$ while the defector gets the temptation $T$. In general, these payoffs must satisfy the ranking $T > R > P \geq S$ and $2R > T + S$, which means that defect is the best strategy from the individual point of view while mutual cooperation may yield higher collective benefit than mutual defection.

Since the original proposition of the prisoner's dilemma, several mechanisms capable of promoting cooperative behavior have received attention. It is worthwhile to mention the seminal idea proposed by May and \text{Nowak} in 1992 \cite{nowak1992evolutionary} which established \text{a} relationship between the topological structure of the game and cooperative behavior. Following this trend, several proposals have investigated the influence of the topological structure on the evolution of cooperative behavior. For instance, games in complex networks, such as random graphs, small-world networks, scale-free networks and multilayer networks, have been intensively studied \cite{abramson2001social,kim2002dynamic,hauert2004spatial,ifti2004effects,santos2005epidemic,santos2005scale,wu2005spatial,tomassini2006hawks, rong2007roles,li2007scale, poncela2007robustness,gomez2007dynamical,szabo2007evolutionary,hatzopoulos2008prisoner, pusch2008impact,chen2008promotion,assenza2008enhancement,yang2009diversity, gomez2012evolution,wang2012probabilistic,gomez2012evolutionary,wang2012evolution,wang2013optimal,wang2013interdependent,jiang2013spreading,santos2014biased,wang2014degree,jin2014spontaneous}. Moreover, other factors have been shown to influence the cooperative behavior (for example, reputation \cite{fu2008reputation}, voluntary participation \cite{szabo2002evolutionary, hauert2002volunteering}, social diversity \cite{perc2008social}, memory \cite{wang2006memory}, mobile agents \cite{meloni2009effects, wu2009effects,jiang2010role}, and many others). 

In this work, we narrow our attention to the possibility that agents get asymmetric payoffs based on some variable according to which they differ \cite{mcavoy2015asymmetric}. In particular, we would like to investigate a particular coevolutionary process--aging--in the temporal evolution of populations in evolutionary games. Although it can not be denied that a population's age structure influences its dynamics \cite{charlesworth1994evolution}, studies taking aging explicitly into account are relatively few \cite{perc2010coevolutionary}. In particular, Szolnoki et al. \cite{szolnoki2009impact} investigated how different aging protocols can influence the evolution of cooperation when agents learn as time goes by. This is done by means of the introduction of an age-dependent rescaling factor that influences the strategy transfer capability. Apart from that, Wang et al. \cite{wang2012cooperation} also investigated how the age structure of the population can interfere in cooperative behavior. They studied an evolutionary spatial game in which cooperation arises when the payoff correlates with the increasing age of players. Both studies concluded that \text{under} certain circumstances, cooperative behavior is enhanced and it even may prevail over selfish behavior. 

In the same year, Wang et al. \cite{wang2012age} explored a variation of these previous ideas. In this case, agents play on a regular network and they can modify their strategies by imitating one of their neighbors. This possibility is influenced by the age of each neighbor and by a factor that determines the extension of the effect in the model. The study concluded that for a specific range of this parameter, older agents are most often imitated and cooperative behavior may prevail.

Despite these results, some aspects \text{relating} to how cooperative behavior can emerge and be maintained in age-structured games still need to be better understood. In particular, the above-cited studies start from the premise that the ability to transfer one's strategy to other players and lead other agents increases with age. However, in recent years, we have seen a significant change in the paradigm of leadership in various social circles. Skills that are developed with age have depreciated and new skills, such as speech style, have been more valued and have begun to characterize modern leaders. Agents of any age are able to massively influence the population and indicate how to dress, what to eat, how to act, etc.  
A good illustration of this is present in \textit{Time} magazine, which annually publishes a list of the one hundred most influential people in the world for the last year. The list for 2018 is the youngest list of all time. In it, $45\%$ are under 40 years old and one of the personalities is only 14 years old \cite{time2018}. Furthermore, Spisak's work \cite{spisak2012general} showed that in time of conflict, society prefers leaders who appear older.  In times of peace, society prefers younger leaders. It is also possible to observe this among groups of primates where leadership does not always belong to the elders \cite{jacobs2008influence,watanabe2008review,FernÃ¡ndez2013}.


Some research questions follow naturally from these observations:  how can this new leadership scenario be replicated in the framework of evolutionary games? Can agents of any age play \text{a leadership role} and propagate their influences? How would this impact the evolution of populations? Moreover, even in these previous studies, all agents die when they reach a threshold age, which is not biologically realistic. How would the emergence and maintenance of cooperative behavior be affected by the introduction of a more natural mechanism of birth/death in game dynamics? 


The purpose of this paper is to answer these important questions. In particular, we use Monte Carlo simulations to show that the strengthening of cooperative behavior depends exclusively on the type of function used in building the age-preference of the game. 
The remainder of this paper is structured as follows.  In the next section, we present the model. Section III is devoted to the presentation of the results.  Lastly, we summarize the paper and make some remarks. 

\section{THE MODEL} 

We consider the  evolutionary prisoner's dilemma in a standard parametrization: the highest payoff received by a defector if playing against a cooperator represents the temptation to defect $1<T < 2$, the punishment for mutual defection is $P = 0$, the sucker's payoff is $S=0$ (weak version of prisoner's dilemma) and the reward for mutual cooperation is $R = 1$. Each player can be either a defector (D) or a cooperator (C). Initially, there \text{are} an equal number of both. The contact network is a regular square lattice. Each player interacts only with their nearest neighbors and we use periodic boundary conditions. The aging process is included as follows: at the beginning, each player is assigned a random integer age $A_i$ between zero and an arbitrary maximum age $A_{max}=50$. At each Monte Carlo step (MCS), all ages are increased by one. Players die with probability defined by a sigmoid distribution:
\begin{equation}
\delta(A)=\frac{1}{1+\alpha \exp(-\beta  A)},
\end{equation}
where $\alpha=16000$ and $\beta=0.1$ are defined to ensure that the average life expectancy is 75 years. There are other ways of ensuring the same life expectancy. However, these ways do not produce the same distribution of ages that we observe in a human population \cite{penna1995bit, chen2003life}. 

When a player dies, their age is reset and they keep their strategy, mimicking a birth-death process. At each round (MCS), all agents play against all their nearest neighbors, receiving the raw payoff $p_i$ (the sum of the payoff resulting from the game with the 4 neighbors) which is \text{then} rescaled by an age-dependent \text{Gaussian} function: 

\begin{equation} \label{resc}
\Pi_i = p_i \left( exp \left(- \frac{(A_i - A_{priv})^2}{2 \sigma^2} \right) \right),
\end{equation}

\noindent
where $A_{priv}$ is the age at which the Gaussian \text{function} is centered and $\sigma$ is the width of the \text{function}. 

In the model, the possibility for agents of any age to exercise leadership corresponds to the free choice of the parameter $A_{priv}$. Besides that, the use of a Gaussian distribution allows us to extend the rescaled payoff benefit to a controlled group of individuals whose age is around $A_{priv}$. The size of this group is determined by the parameter $\sigma$. The granting of benefits to individuals whose age is near to $A_{priv}$ is based on empirical evidence of the existence of an age similarity preference (ASP) in groups of individuals that interact with a specific goal \cite{avery2007engaging}. In contrast, studies indicate that groups of individuals with dissimilar experiences, beliefs, and values (as a result, for example, of age differences) may have communication and social integration difficulties \cite{wiersema1992top, van2002effects}. Furthermore, if $\sigma$ is small, Eq.~\ref{resc} introduces asymmetry to the game and some ages around $A_{priv}$ are privileged. On the other hand, if the value of $\sigma$ is too large ($\sigma\rightarrow \infty$), this asymmetry is diminished because too many agents would benefit by the rescaling process, such that the payoff differences among the majority of players will become very small and leadership will disappear. 

{Humans tend to imitate other humans recognized as leaders \cite{maccoby2004people,braha2017voting}. The model reproduces this possibility by allowing each agent to update their strategy by copying the behavior of another agent they played with in the previous round. Accordingly, one of its four neighbors, whose rescaled payoff is $\Pi_j$, is selected at random. Then, player $i$ adopts the strategy $\phi_j$ from the selected player $j$ with the probability}

\begin{equation}
\label{transition}
W(\phi _i \rightarrow \phi_j) = \frac{1}{1+ \exp[(\Pi_i - \Pi_j)/K]},
\end{equation}

\noindent
where $1/K$ is a measure of noise \cite{szabo1998evolutionary}. { If $\Pi_i$ is equal to $\Pi_j$, the player adopts the strategy of its neighbor with a 50\% chance. However, if $\Pi_j$ is bigger than $\Pi_i$, imitation is more probable. The only factor that determines if imitation occurs is the difference between payoffs. That is why equation (\ref{transition}) only took into account the rescaled payoffs of the players. Obviously, if $\sigma=0$ the game becomes random because for all individuals $\Pi_i=0$ and consequently Eq. (\ref{transition}) always returns $1/2$.}

Next, we present the results of Monte Carlo simulations obtained for lattices with $100^2$ to $400^2$ agents. The average fraction of cooperators (defectors) $\rho_C$ ($\rho_D$) is the number of cooperators (defectors) divided by total number of agents. Our results were obtained within the last $10^4$ out of the total $2 \cdot 10^5$ MCS. Furthermore, the final results were averaged over $40$ independent runs for each set of parameter values.
\section{RESULTS}
\begin{figure}[ht] 
	\centering
	\includegraphics[scale=0.35]{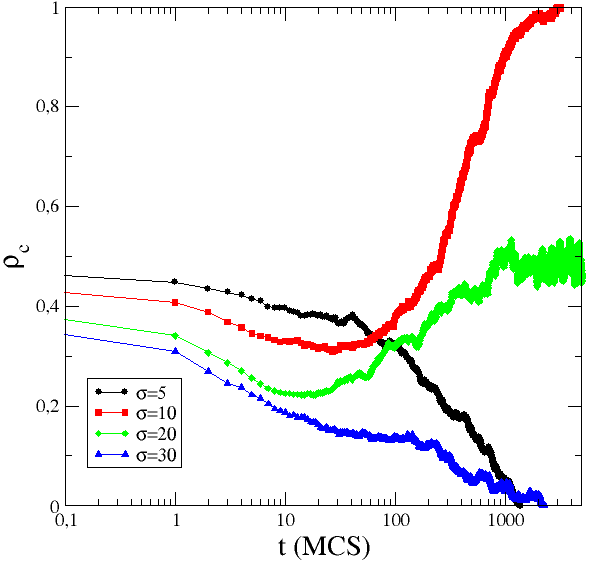} 
	\caption{Time evolution of the \text{density} of cooperators for different values of $\sigma$, $T = 1.1$, $S = 0$, and $K = 0.1$. Since the initial state is random and the cooperators are not yet organized in compact groups, many of them are easy prey for defectors and there always is an initial decrease \text{in} their number. For a given value of $A_{priv}$, values of sigma that are too small ($\sigma$ = 5) or too large ($\sigma$ = 30) mean that, over time, the population of cooperators will decrease and finally be extinguished. \text{For} intermediate values of $\sigma$ ($\sigma$ = $10$ and $20$), the initial decay of the population occurs more slowly, which is due to the greater resistance of cooperator clusters against invasion. In these cases, they can coexist or even dominate the population of defectors.}
	\label{figura1}
\end{figure}

In Fig.\ref{figura1}, we present the temporal evolution of the density of cooperators, $\rho_C(t)$, for several values of $\sigma$, $A_{priv} = 25$ and $T = 1.1$ to help us understand the effect of using a Gaussian distribution to rescale the payoff  in promoting the cooperation. Initially, in all cases, the average \text{density} of cooperators \text{decreases} quickly, which happens because cooperators still \text{are not} organized in compact clusters and they are easily invaded by defectors.  In fact, since the initial state is randomly chosen, it is expected that defectors are more successful in the beginning. Nevertheless, different values of $\sigma$ result in diverse trends at large times. On the one hand, if $\sigma=0$, all payoffs are rescaled to zero and the game becomes a random movement between cooperation and defection. On the other hand, if $\sigma \neq 0$, the outcome of the game is strongly influenced by the values of $\sigma$. If $\sigma =5$, a very small fraction \text{of the} population is benefited by the rescaling process of the payoff, and cooperator clusters are not strong enough. Consequently, over time, the cooperative population disappears. Similarly, for values of $\sigma$ that are too large (e.g., 30), cooperators do not survive \text{either}, though they take a long time to die out. In this case, a too large \text{proportion} of the population is benefited by the rescaling process, the leaders disappear, and consequently the clusters become smaller and more numerous. For intermediate values of $\sigma$ the initial decay of cooperators is stopped and it is followed by a rapid spread of cooperators, until they eventually cover the complete domain of the population or until a stable equilibrium is reached between cooperators and defectors, as shown for $\sigma = 10$ and $\sigma = 20$. Nonetheless, for a given $A_{priv}$, there is an optimal value for \text{the} distribution width ($\sigma$) that enhances cooperative behavior, so that cooperators can not only survive but dominate the population.

\begin{figure}[h] 
	\centering
	\includegraphics[scale=0.295]{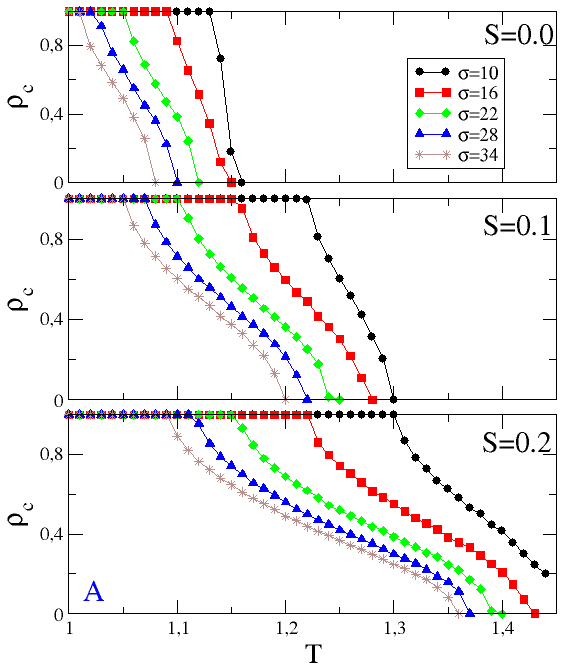} 
	\includegraphics[scale=0.275]{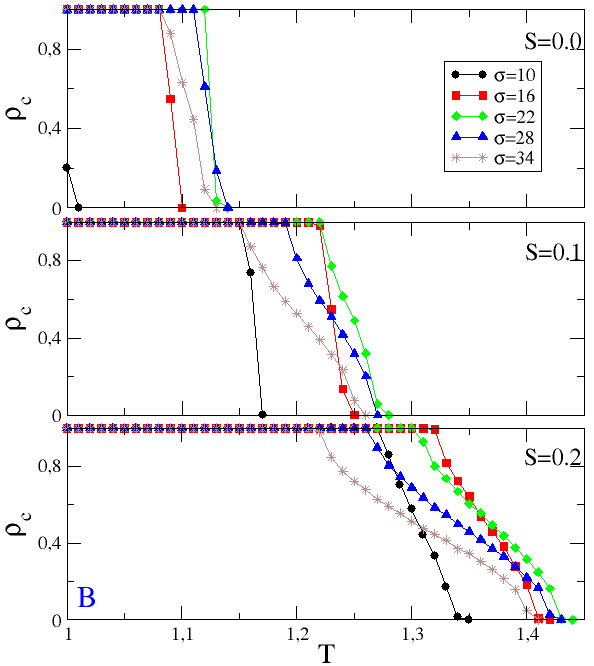} 
	\caption{Asymptotic \text{proportion} of cooperators $\rho_C$ as a function of the temptation $T$ for different values of $\sigma$ and $S$, with $K = 0.1$ and with $A_{priv} = 25$ in (a) and $A_{priv} = 75$ in (b). As expected, higher values of $S$ 
	facilitate the cooperation. For small values of $T$, for each $A_{priv}$,  cooperators dominate the system and there is an optimum $\sigma$ for which cooperation is enhanced. It is justified by the use of a sigmoid distribution (a biologically valid way) to determine the death in the game. As $A_{priv}$ increases, the optimal value for $\sigma$ to promote the cooperation needs to be higher because it becomes less and less likely to encounter old agents.  
	}
	\label{figura2}
\end{figure}

In Fig. \ref{figura2} we show the steady state of the system for $A_{priv}=25$ (\ref{figura2}(a)) and $A_{priv}=75$ (\ref{figura2}(b)) with different values of $\sigma$, temptation ($T$) and sucker's payoff ($S$). Moreover, \text{we let $T_C$ denote} the maximum value of \text{the temptation payoff} where all players becomes cooperators and \text{we let $T_D$ denote} the minimum value of temptation where all players become defectors. As we can see in Fig. \ref{figura2}(a), $T_C$ and $T_D$ depend on $\sigma$. For example, for $S=0$ when $\sigma=10$, for $T_C =1.13$, $T_D=1.16$ and for $1.13<T<1.16$ the system goes to a \text{state of} coexistence between cooperators and defectors. Similarly, we can see in Fig. \ref{figura2}(b) that $T_C$ and $T_D$ also \text{change} with $\sigma$. \text{These results strongly suggest} that, for given $A_{priv}$, there is \text{an} optimal $\sigma$ that \text{promotes} cooperation.

\begin{figure}[ht] 
	\centering
	\includegraphics[scale=0.22]{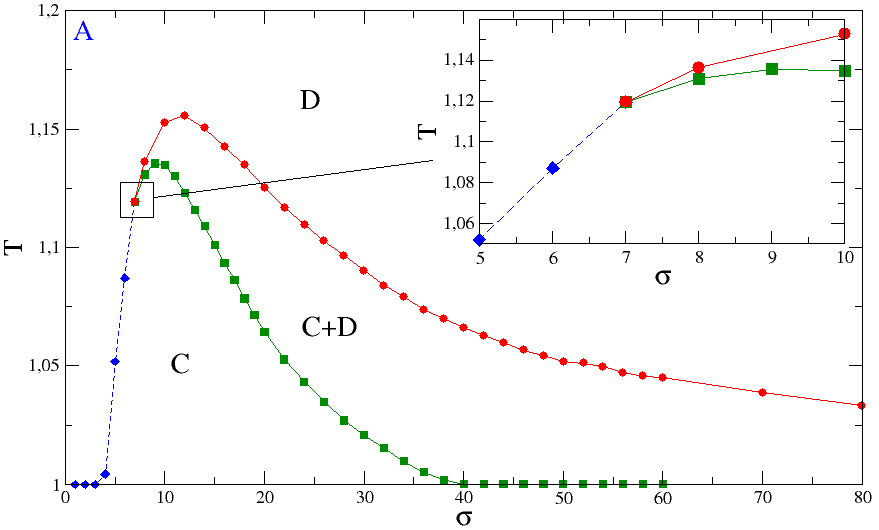} \\ \includegraphics[scale=0.22]{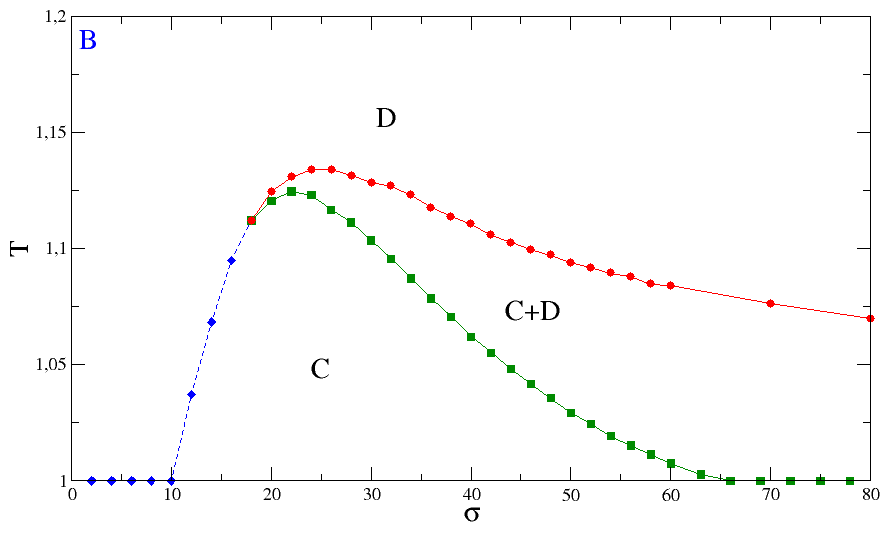} 
	\caption{Phase diagram $T$ $\&$ $\sigma$. In (a), $A_{priv} = 25$ while in (b), $A_{priv} = 75$. The red (green) line represents the critical threshold above (below) which cooperators (defectors) go extinct. In the region between the lines, there is a coexistence of strategies. The blue line corresponds to a first order transition between two absorvent states (full cooperation or full defection). The optimal value of $\sigma$ to promote cooperative behavior depends on  $A_{priv}$, as does the saturation value. As $A_{priv}$ increases, the graph is shifted to the right.}
	\label{figura3}
\end{figure}

The existence of an optimal $\sigma$ for a given value of $A_{priv}$ becomes clearer in the phase diagram shown in Fig.\ref{figura3}. In Fig. \ref{figura3}(a), $A_{priv} = 25$ and in Fig. \ref{figura3}(b), $A_{priv} = 75$. The green line represents $T_C(\sigma)$, which means that under this line, independent of $T$ and $\sigma$, the system goes to a full cooperation state (C). The red line represents $T_D$, which means that above this line the system goes to a full defection state (D). Between these lines, the system goes to coexistence of cooperators and defectors (C+D). The blue line represents the curve where $T_C=T_D$, which means that it is impossible to observe the system in a coexistence state. The transition between full cooperation and full defection is not continuous. Our model can be classified as a contact process (CP) which belongs to the directed percolation universality class (DP). However, the DP class has a continuous transition. In 1974, A. B. Harris \cite{harris} proposed that the universality class of some systems can be changed by the presence of random defects in the lattice. The Harris's results are reproduced in CP problems \cite{dickman, wagner} where some impurities are introduced to the system. In Fig \ref{figura3} we show that the change in universality class only occurs for small values of $\sigma$. We do not have introduced impurities or defects in our system but, when $\sigma$ is very small, the payoff is always adjusted to roughly zero for individuals whose age is far away from the $A_{priv}$. In our model, individuals with zero payoffs, do not influence other individuals and act as a defect in the lattice. The inset in Fig. \ref{figura3}(a) depicts the exact point where $T_C\neq T_D$. This point is determined by a finite size scaling method. The overall behavior of these graphs is quite similar. The higher the privileged age ($A_{priv}$, the age at which the Gaussian is centered) is, the higher the optimal $\sigma$ value that enhances cooperative behavior is. This is due to the following fact: since agents die according to a sigmoid distribution, fewer and fewer agents remain alive as time goes by. Thus, for older agents to be able to lead the way in the formation of clusters of cooperators, the value of $\sigma$ that optimizes cooperative behavior is higher than when the privileged age is younger. Nonetheless, for a given privileged age, there must be a gradient of improvement that can neither be too abrupt nor too smooth for cooperative behavior to be enhanced.

\begin{figure}[H] 
	\centering
	\includegraphics[scale=0.6]{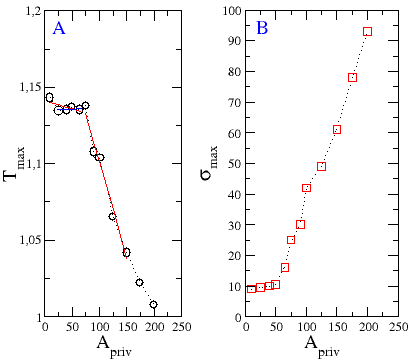}
	\caption{(A) presents the maximum value of
Temptation ($T$) where the cooperators dominate the population as a function of $A_{priv}$. (B) presents the maximum value of $\sigma$ wherein the cooperators can dominate the population as a function of $A_{priv}$. The behavior observed in both graphs are immediate consequences of the use of a sigmoid distribution to simulate the process of birth and death.}
	\label{figura6}
\end{figure}


In Fig. \ref{figura6} we show the graphs of $T_{max}$ $\&$ $A_{priv}$ (a) and $\sigma_{max}$ $\&$ $A_{priv}$ (b). $T_{max}$ ($\sigma_{max}$) is the maximum value of 
T($\sigma$) where the cooperators dominate the population. Since the \text{tenacity of cooperative behavior} depends on the leadership exercised by agents whose age is privileged (near $A_{priv}$), the behavior observed in Fig. \ref{figura6}(a) is due to the fact that the birth/death device is modeled by a sigmoid function. For this reason, the payoff rescheduling process becomes increasingly inefficient as $A_{priv}$ grows, especially when the $A_{priv}$ \text{exceeds} 100 years and the value of $T_{max}$ strongly decreases. Similarly, as $A_{priv}$ increases, the optimal Gaussian width ($\sigma$) must be increased for the cooperators to maintain dominance over the population \text{for} the same reason explained before, as shown in Fig. \ref{figura6}(b).

\begin{figure}[!ht] 
	\centering
	\includegraphics[scale=0.25]{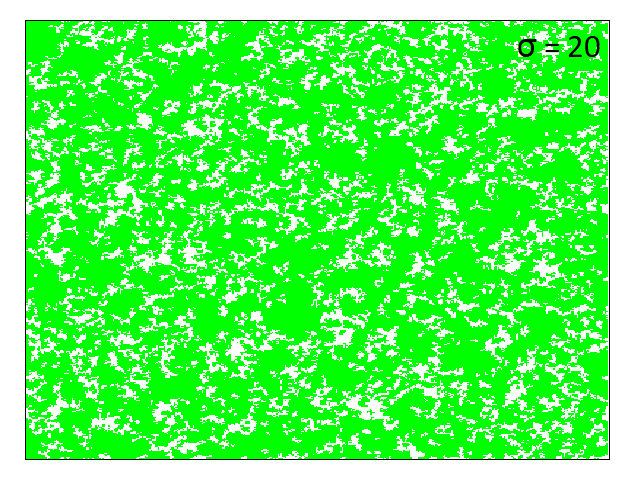}  \includegraphics[scale=0.25]{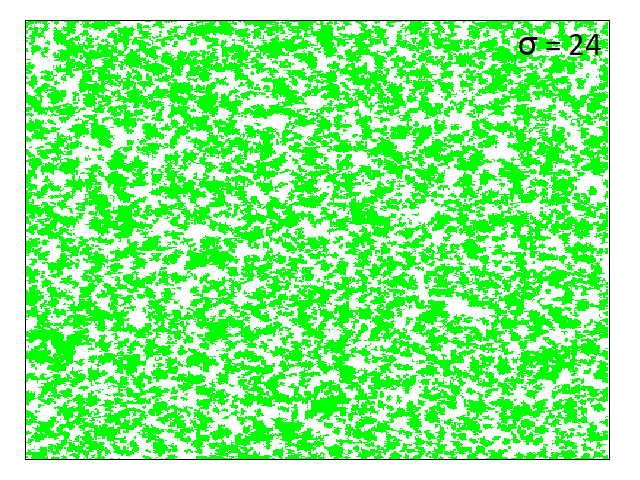} \\ \includegraphics[scale=0.25]{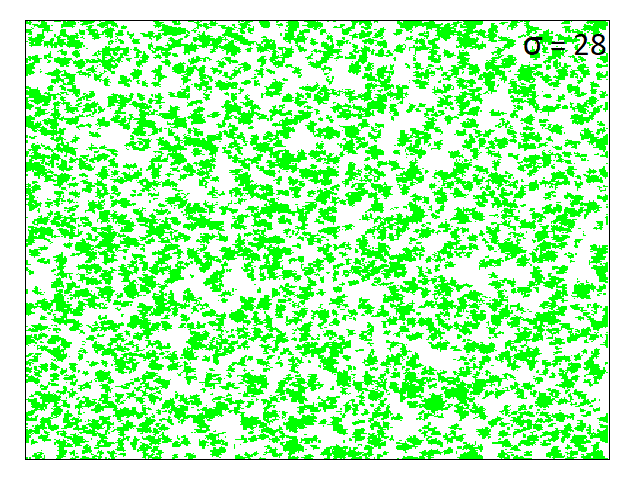}  \includegraphics[scale=0.25]{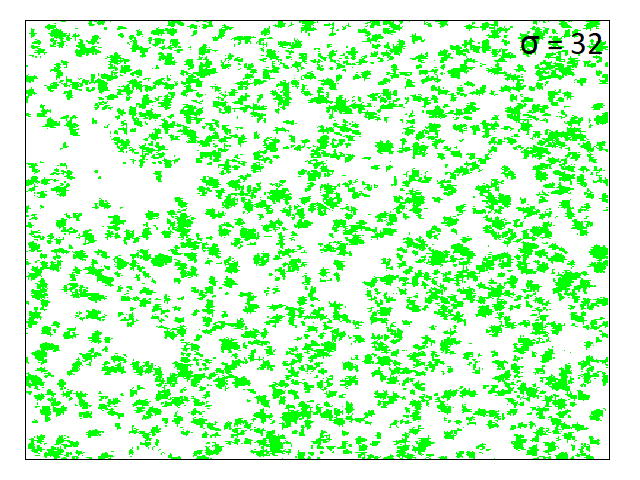}
	\caption{Lattice snapshots showing the clusters of cooperators on a $400^2$ square lattice with $S = 0$, $T = 1.1$, $A_{priv} = 75$ and $K = 0.1$ for $\sigma =20, 24, 28,$ and $32$ after $10^5$ MCS. The number of clusters evidently decreases as $\sigma$ increases after reaching its optimal value.}
	\label{figura4}
\end{figure}

Moreover, it is instructive to analyze typical spatial configurations of cooperators and defectors for different values of $\sigma$, as shown in Fig. \ref{figura4} for $T = 1.1$, $A_{priv} = 75$ and $K = 0.1$. Cooperators are represented in green. They tend to cluster in \text{``islands"} of cooperators to resist ``invasion" \text{by} defectors. These clusters are formed around players whose age is \text{in the neighborhood of} $A_{priv}$. 
If $\sigma$ is not large enough, the players from the edge of the cluster, who are necessarily much younger (or older) than the leaders, are easily attracted to defection because it is a more profitable behavior \text{played} against defectors. \text{This} means that for small $\sigma$ the clustering strength is weak. Furthermore, if $\sigma$ is too large, all players have almost the same rescaling factor, and cooperation loses to defection equally. On the other hand, if $\sigma$ presents an intermediate optimal value, cooperators \text{do} not only survive but dominate the population. 

\begin{figure}[!ht] 
	\centering
	\includegraphics[scale=0.27]{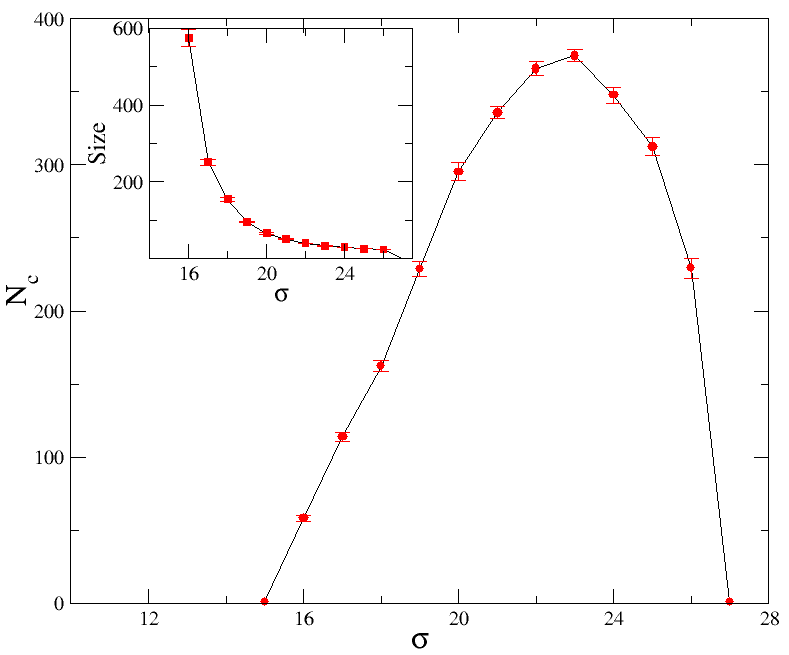} 
	\caption{ Stationary number of clusters of cooperators $(N_C)$ for a $400^2$ lattice with $S = 0$, $T = 1.1$, $A_{priv} = 75$ and $K = 0.1$ as a function of $\sigma$. In the inset, the stationary average size of clusters of cooperators also as a function of $\sigma$ is shown. There is a specific value for $\sigma$ that maximizes the number of clusters. In addition, the size of these clusters decreases as $\sigma$ increases.}
	\label{figura5}
\end{figure}

Fig. \ref{figura5} shows the number of clusters of cooperators ($N_C$) and $\sigma$ for $A_{priv}=75$. At $\sigma=15$ all players are cooperators (see figure \ref{figura3}(b)), so $N_C=1$ because all players belong to the same cluster. As $\sigma$ increases, the system goes to a coexistence phase and $N_C$ grows as the defectors \text{invade} the clusters. The number of clusters keeps growing up to $\sigma=23$ and after, decreases to $N_C=0$ because all players become defectors. The inset in Fig. \ref{figura5} shows the mean size (number of players) of the clusters \textit{versus} $\sigma$. As defectors \text{break} the clusters, the size decreases \text{until it reaches zero}.

Since the payoff rescaling affects all agents according to their age, how does this process benefit cooperation? Improving the payoff of some agents facilitates the formation of clusters of players who share the same strategy and are led by the players who benefit from the rescaling process, both cooperators and defectors. But in the prisoner's dilemma, mutual cooperation results in \text{a} bigger payoff than mutual defection, which makes cooperator clusters more stable than defector clusters when an optimum range of players, dependent on the age at which the Gaussian distribution is centered, is benefited by the process. 

\section{FINAL REMARKS}

We considered an age-structured prisoner's dilemma game where the payoff is \text{a function of age}. 
In this game, although the raw payoff \text{for} each agent is a consequence of \text{their} strategy as well as the strategies of the other agents with whom \text{they interact}, a Gaussian distribution is used to modify the payoff obtained each round. In this case, the age at which the Gaussian is centered and its width are controllable parameters and simulate the ability of agents of any age to exert leadership and influence other agents to utilize the same strategy of play. In addition, a sigmoid distribution is used in the simulation of the birth and death process of the agents, which makes the game more realistic and has a profound influence on its outcome. \\

The results indicate that each privileged age in the game (\text{the} age at which the Gaussian \text{function} is centered) is associated with a range of widths in which agents adopting cooperation as the strategy may not only survive but dominate the population. 
As previously indicated in the literature, the promotion and enhancement of cooperative behavior are attributed to the formation of clusters of cooperators, which allows cooperators to resist the invasion of the defectors \cite{szolnoki2009impact,wang2012cooperation}. 
This is obviously a consequence of mutual cooperation being more attractive than mutual betrayal in a PD game. In fact, once \text{the privileged age has been specified}, if the width of the Gaussian is smaller than a minimum value, these clusters are not \text{sufficiently robust} and the cooperator \text{clusters} are invaded by defectors. However, if \text{the width} is larger than a maximum value, too many agents are benefited, leadership disappears and the number and the size of \text{cooperator clusters decreases}. This dynamic is a consequence of the age structure \text{and the birth and death processes} of the game. \\

Although these results help in understanding how age can influence coevolutionary processes, there are still some issues that need to be better understood, such as how system dynamics are affected when more exotic distributions are used to implement the age improvement or how these dynamics are modified when the interaction topology is more realistically modeled through complex networks.

\section{ACKNOWLEDGMENTS}

We would like to thank the Brazilian agencies CNPq and CAPES for their financial support of this research. We also thank Eric Kuo for useful comments.
\bibliography{TDJREF}

\begin{thebibliography}{62}%
\makeatletter
\providecommand \@ifxundefined [1]{%
 \@ifx{#1\undefined}
}%
\providecommand \@ifnum [1]{%
 \ifnum #1\expandafter \@firstoftwo
 \else \expandafter \@secondoftwo
 \fi
}%
\providecommand \@ifx [1]{%
 \ifx #1\expandafter \@firstoftwo
 \else \expandafter \@secondoftwo
 \fi
}%
\providecommand \natexlab [1]{#1}%
\providecommand \enquote  [1]{``#1''}%
\providecommand \bibnamefont  [1]{#1}%
\providecommand \bibfnamefont [1]{#1}%
\providecommand \citenamefont [1]{#1}%
\providecommand \href@noop [0]{\@secondoftwo}%
\providecommand \href [0]{\begingroup \@sanitize@url \@href}%
\providecommand \@href[1]{\@@startlink{#1}\@@href}%
\providecommand \@@href[1]{\endgroup#1\@@endlink}%
\providecommand \@sanitize@url [0]{\catcode `\\12\catcode `\$12\catcode
  `\&12\catcode `\#12\catcode `\^12\catcode `\_12\catcode `\%12\relax}%
\providecommand \@@startlink[1]{}%
\providecommand \@@endlink[0]{}%
\providecommand \url  [0]{\begingroup\@sanitize@url \@url }%
\providecommand \@url [1]{\endgroup\@href {#1}{\urlprefix }}%
\providecommand \urlprefix  [0]{URL }%
\providecommand \Eprint [0]{\href }%
\providecommand \doibase [0]{http://dx.doi.org/}%
\providecommand \selectlanguage [0]{\@gobble}%
\providecommand \bibinfo  [0]{\@secondoftwo}%
\providecommand \bibfield  [0]{\@secondoftwo}%
\providecommand \translation [1]{[#1]}%
\providecommand \BibitemOpen [0]{}%
\providecommand \bibitemStop [0]{}%
\providecommand \bibitemNoStop [0]{.\EOS\space}%
\providecommand \EOS [0]{\spacefactor3000\relax}%
\providecommand \BibitemShut  [1]{\csname bibitem#1\endcsname}%
\let\auto@bib@innerbib\@empty
\bibitem [{\citenamefont {Nowak}(2006)}]{nowak2006five}%
  \BibitemOpen
  \bibfield  {author} {\bibinfo {author} {\bibfnamefont {M.~A.}\ \bibnamefont
  {Nowak}},\ }\href@noop {} {\bibfield  {journal} {\bibinfo  {journal}
  {Science}\ }\textbf {\bibinfo {volume} {314}},\ \bibinfo {pages} {1560}
  (\bibinfo {year} {2006})}\BibitemShut {NoStop}%
\bibitem [{\citenamefont {Perc}\ \emph {et~al.}(2017)\citenamefont {Perc},
  \citenamefont {Jordan}, \citenamefont {Rand}, \citenamefont {Wang},
  \citenamefont {Boccaletti},\ and\ \citenamefont
  {Szolnoki}}]{perc2017statistical}%
  \BibitemOpen
  \bibfield  {author} {\bibinfo {author} {\bibfnamefont {M.}~\bibnamefont
  {Perc}}, \bibinfo {author} {\bibfnamefont {J.~J.}\ \bibnamefont {Jordan}},
  \bibinfo {author} {\bibfnamefont {D.~G.}\ \bibnamefont {Rand}}, \bibinfo
  {author} {\bibfnamefont {Z.}~\bibnamefont {Wang}}, \bibinfo {author}
  {\bibfnamefont {S.}~\bibnamefont {Boccaletti}}, \ and\ \bibinfo {author}
  {\bibfnamefont {A.}~\bibnamefont {Szolnoki}},\ }\href@noop {} {\bibfield
  {journal} {\bibinfo  {journal} {{P}hysics {R}eports}\ }\textbf {\bibinfo
  {volume} {687}},\ \bibinfo {pages} {1} (\bibinfo {year} {2017})}\BibitemShut
  {NoStop}%
\bibitem [{\citenamefont {Smith}\ and\ \citenamefont
  {Price}(1973)}]{smith1973lhe}%
  \BibitemOpen
  \bibfield  {author} {\bibinfo {author} {\bibfnamefont {J.~M.}\ \bibnamefont
  {Smith}}\ and\ \bibinfo {author} {\bibfnamefont {G.~R.}\ \bibnamefont
  {Price}},\ }\href@noop {} {\bibfield  {journal} {\bibinfo  {journal}
  {Nature}\ }\textbf {\bibinfo {volume} {246}},\ \bibinfo {pages} {15}
  (\bibinfo {year} {1973})}\BibitemShut {NoStop}%
\bibitem [{\citenamefont {Rapoport}\ \emph {et~al.}(1965)\citenamefont
  {Rapoport}, \citenamefont {Chammah},\ and\ \citenamefont
  {Orwant}}]{rapoport1965prisoner}%
  \BibitemOpen
  \bibfield  {author} {\bibinfo {author} {\bibfnamefont {A.}~\bibnamefont
  {Rapoport}}, \bibinfo {author} {\bibfnamefont {A.~M.}\ \bibnamefont
  {Chammah}}, \ and\ \bibinfo {author} {\bibfnamefont {C.~J.}\ \bibnamefont
  {Orwant}},\ }\href@noop {} {\emph {\bibinfo {title} {Prisoner's dilemma: A
  study in conflict and cooperation}}},\ Vol.\ \bibinfo {volume} {165}\
  (\bibinfo  {publisher} {University of Michigan Press},\ \bibinfo {year}
  {1965})\BibitemShut {NoStop}%
\bibitem [{\citenamefont {Nowak}\ and\ \citenamefont
  {May}(1992)}]{nowak1992evolutionary}%
  \BibitemOpen
  \bibfield  {author} {\bibinfo {author} {\bibfnamefont {M.~A.}\ \bibnamefont
  {Nowak}}\ and\ \bibinfo {author} {\bibfnamefont {R.~M.}\ \bibnamefont
  {May}},\ }\href@noop {} {\bibfield  {journal} {\bibinfo  {journal} {Nature}\
  }\textbf {\bibinfo {volume} {359}},\ \bibinfo {pages} {826} (\bibinfo {year}
  {1992})}\BibitemShut {NoStop}%
\bibitem [{\citenamefont {Abramson}\ and\ \citenamefont
  {Kuperman}(2001)}]{abramson2001social}%
  \BibitemOpen
  \bibfield  {author} {\bibinfo {author} {\bibfnamefont {G.}~\bibnamefont
  {Abramson}}\ and\ \bibinfo {author} {\bibfnamefont {M.}~\bibnamefont
  {Kuperman}},\ }\href@noop {} {\bibfield  {journal} {\bibinfo  {journal}
  {Physical Review E}\ }\textbf {\bibinfo {volume} {63}},\ \bibinfo {pages}
  {030901} (\bibinfo {year} {2001})}\BibitemShut {NoStop}%
\bibitem [{\citenamefont {Kim}\ \emph {et~al.}(2002)\citenamefont {Kim},
  \citenamefont {Trusina}, \citenamefont {Holme}, \citenamefont {Minnhagen},
  \citenamefont {Chung},\ and\ \citenamefont {Choi}}]{kim2002dynamic}%
  \BibitemOpen
  \bibfield  {author} {\bibinfo {author} {\bibfnamefont {B.~J.}\ \bibnamefont
  {Kim}}, \bibinfo {author} {\bibfnamefont {A.}~\bibnamefont {Trusina}},
  \bibinfo {author} {\bibfnamefont {P.}~\bibnamefont {Holme}}, \bibinfo
  {author} {\bibfnamefont {P.}~\bibnamefont {Minnhagen}}, \bibinfo {author}
  {\bibfnamefont {J.~S.}\ \bibnamefont {Chung}}, \ and\ \bibinfo {author}
  {\bibfnamefont {M.}~\bibnamefont {Choi}},\ }\href@noop {} {\bibfield
  {journal} {\bibinfo  {journal} {Physical Review E}\ }\textbf {\bibinfo
  {volume} {66}},\ \bibinfo {pages} {021907} (\bibinfo {year}
  {2002})}\BibitemShut {NoStop}%
\bibitem [{\citenamefont {Hauert}\ and\ \citenamefont
  {Doebeli}(2004)}]{hauert2004spatial}%
  \BibitemOpen
  \bibfield  {author} {\bibinfo {author} {\bibfnamefont {C.}~\bibnamefont
  {Hauert}}\ and\ \bibinfo {author} {\bibfnamefont {M.}~\bibnamefont
  {Doebeli}},\ }\href@noop {} {\bibfield  {journal} {\bibinfo  {journal}
  {Nature}\ }\textbf {\bibinfo {volume} {428}},\ \bibinfo {pages} {643}
  (\bibinfo {year} {2004})}\BibitemShut {NoStop}%
\bibitem [{\citenamefont {Ifti}\ \emph {et~al.}(2004)\citenamefont {Ifti},
  \citenamefont {Killingback},\ and\ \citenamefont
  {Doebeli}}]{ifti2004effects}%
  \BibitemOpen
  \bibfield  {author} {\bibinfo {author} {\bibfnamefont {M.}~\bibnamefont
  {Ifti}}, \bibinfo {author} {\bibfnamefont {T.}~\bibnamefont {Killingback}}, \
  and\ \bibinfo {author} {\bibfnamefont {M.}~\bibnamefont {Doebeli}},\
  }\href@noop {} {\bibfield  {journal} {\bibinfo  {journal} {Journal of
  {T}heoretical {B}iology}\ }\textbf {\bibinfo {volume} {231}},\ \bibinfo
  {pages} {97} (\bibinfo {year} {2004})}\BibitemShut {NoStop}%
\bibitem [{\citenamefont {Santos}\ \emph {et~al.}(2005)\citenamefont {Santos},
  \citenamefont {Rodrigues},\ and\ \citenamefont
  {Pacheco}}]{santos2005epidemic}%
  \BibitemOpen
  \bibfield  {author} {\bibinfo {author} {\bibfnamefont {F.~C.}\ \bibnamefont
  {Santos}}, \bibinfo {author} {\bibfnamefont {J.~F.}\ \bibnamefont
  {Rodrigues}}, \ and\ \bibinfo {author} {\bibfnamefont {J.~M.}\ \bibnamefont
  {Pacheco}},\ }\href@noop {} {\bibfield  {journal} {\bibinfo  {journal}
  {Physical Review E}\ }\textbf {\bibinfo {volume} {72}},\ \bibinfo {pages}
  {056128} (\bibinfo {year} {2005})}\BibitemShut {NoStop}%
\bibitem [{\citenamefont {Santos}\ and\ \citenamefont
  {Pacheco}(2005)}]{santos2005scale}%
  \BibitemOpen
  \bibfield  {author} {\bibinfo {author} {\bibfnamefont {F.~C.}\ \bibnamefont
  {Santos}}\ and\ \bibinfo {author} {\bibfnamefont {J.~M.}\ \bibnamefont
  {Pacheco}},\ }\href@noop {} {\bibfield  {journal} {\bibinfo  {journal}
  {Physical {R}eview {L}etters}\ }\textbf {\bibinfo {volume} {95}},\ \bibinfo
  {pages} {098104} (\bibinfo {year} {2005})}\BibitemShut {NoStop}%
\bibitem [{\citenamefont {Wu}\ \emph {et~al.}(2005)\citenamefont {Wu},
  \citenamefont {Xu}, \citenamefont {Chen},\ and\ \citenamefont
  {Wang}}]{wu2005spatial}%
  \BibitemOpen
  \bibfield  {author} {\bibinfo {author} {\bibfnamefont {Z.-X.}\ \bibnamefont
  {Wu}}, \bibinfo {author} {\bibfnamefont {X.-J.}\ \bibnamefont {Xu}}, \bibinfo
  {author} {\bibfnamefont {Y.}~\bibnamefont {Chen}}, \ and\ \bibinfo {author}
  {\bibfnamefont {Y.-H.}\ \bibnamefont {Wang}},\ }\href@noop {} {\bibfield
  {journal} {\bibinfo  {journal} {Physical Review E}\ }\textbf {\bibinfo
  {volume} {71}},\ \bibinfo {pages} {037103} (\bibinfo {year}
  {2005})}\BibitemShut {NoStop}%
\bibitem [{\citenamefont {Tomassini}\ \emph {et~al.}(2006)\citenamefont
  {Tomassini}, \citenamefont {Luthi},\ and\ \citenamefont
  {Giacobini}}]{tomassini2006hawks}%
  \BibitemOpen
  \bibfield  {author} {\bibinfo {author} {\bibfnamefont {M.}~\bibnamefont
  {Tomassini}}, \bibinfo {author} {\bibfnamefont {L.}~\bibnamefont {Luthi}}, \
  and\ \bibinfo {author} {\bibfnamefont {M.}~\bibnamefont {Giacobini}},\
  }\href@noop {} {\bibfield  {journal} {\bibinfo  {journal} {Physical Review
  E}\ }\textbf {\bibinfo {volume} {73}},\ \bibinfo {pages} {016132} (\bibinfo
  {year} {2006})}\BibitemShut {NoStop}%
\bibitem [{\citenamefont {Rong}\ \emph {et~al.}(2007)\citenamefont {Rong},
  \citenamefont {Li},\ and\ \citenamefont {Wang}}]{rong2007roles}%
  \BibitemOpen
  \bibfield  {author} {\bibinfo {author} {\bibfnamefont {Z.}~\bibnamefont
  {Rong}}, \bibinfo {author} {\bibfnamefont {X.}~\bibnamefont {Li}}, \ and\
  \bibinfo {author} {\bibfnamefont {X.}~\bibnamefont {Wang}},\ }\href@noop {}
  {\bibfield  {journal} {\bibinfo  {journal} {Physical Review E}\ }\textbf
  {\bibinfo {volume} {76}},\ \bibinfo {pages} {027101} (\bibinfo {year}
  {2007})}\BibitemShut {NoStop}%
\bibitem [{\citenamefont {Li}\ \emph {et~al.}(2007)\citenamefont {Li},
  \citenamefont {Zhang},\ and\ \citenamefont {Hu}}]{li2007scale}%
  \BibitemOpen
  \bibfield  {author} {\bibinfo {author} {\bibfnamefont {W.}~\bibnamefont
  {Li}}, \bibinfo {author} {\bibfnamefont {X.}~\bibnamefont {Zhang}}, \ and\
  \bibinfo {author} {\bibfnamefont {G.}~\bibnamefont {Hu}},\ }\href@noop {}
  {\bibfield  {journal} {\bibinfo  {journal} {Physical Review E}\ }\textbf
  {\bibinfo {volume} {76}},\ \bibinfo {pages} {045102} (\bibinfo {year}
  {2007})}\BibitemShut {NoStop}%
\bibitem [{\citenamefont {Poncela}\ \emph {et~al.}(2007)\citenamefont
  {Poncela}, \citenamefont {G{\'o}mez-Gardenes}, \citenamefont {Flor{\'\i}a},\
  and\ \citenamefont {Moreno}}]{poncela2007robustness}%
  \BibitemOpen
  \bibfield  {author} {\bibinfo {author} {\bibfnamefont {J.}~\bibnamefont
  {Poncela}}, \bibinfo {author} {\bibfnamefont {J.}~\bibnamefont
  {G{\'o}mez-Gardenes}}, \bibinfo {author} {\bibfnamefont {L.~M.}\ \bibnamefont
  {Flor{\'\i}a}}, \ and\ \bibinfo {author} {\bibfnamefont {Y.}~\bibnamefont
  {Moreno}},\ }\href@noop {} {\bibfield  {journal} {\bibinfo  {journal} {New
  {J}ournal of {P}hysics}\ }\textbf {\bibinfo {volume} {9}},\ \bibinfo {pages}
  {184} (\bibinfo {year} {2007})}\BibitemShut {NoStop}%
\bibitem [{\citenamefont {G{\'o}mez-Gardenes}\ \emph
  {et~al.}(2007)\citenamefont {G{\'o}mez-Gardenes}, \citenamefont {Campillo},
  \citenamefont {Flor{\'\i}a},\ and\ \citenamefont
  {Moreno}}]{gomez2007dynamical}%
  \BibitemOpen
  \bibfield  {author} {\bibinfo {author} {\bibfnamefont {J.}~\bibnamefont
  {G{\'o}mez-Gardenes}}, \bibinfo {author} {\bibfnamefont {M.}~\bibnamefont
  {Campillo}}, \bibinfo {author} {\bibfnamefont {L.~M.}\ \bibnamefont
  {Flor{\'\i}a}}, \ and\ \bibinfo {author} {\bibfnamefont {Y.}~\bibnamefont
  {Moreno}},\ }\href@noop {} {\bibfield  {journal} {\bibinfo  {journal}
  {Physical Review Letters}\ }\textbf {\bibinfo {volume} {98}},\ \bibinfo
  {pages} {108103} (\bibinfo {year} {2007})}\BibitemShut {NoStop}%
\bibitem [{\citenamefont {Szab{\'o}}\ and\ \citenamefont
  {Fath}(2007)}]{szabo2007evolutionary}%
  \BibitemOpen
  \bibfield  {author} {\bibinfo {author} {\bibfnamefont {G.}~\bibnamefont
  {Szab{\'o}}}\ and\ \bibinfo {author} {\bibfnamefont {G.}~\bibnamefont
  {Fath}},\ }\href@noop {} {\bibfield  {journal} {\bibinfo  {journal} {Physics
  {R}eports - {N}ature}\ }\textbf {\bibinfo {volume} {446}},\ \bibinfo {pages}
  {97} (\bibinfo {year} {2007})}\BibitemShut {NoStop}%
\bibitem [{\citenamefont {Hatzopoulos}\ and\ \citenamefont
  {Jensen}(2008)}]{hatzopoulos2008prisoner}%
  \BibitemOpen
  \bibfield  {author} {\bibinfo {author} {\bibfnamefont {V.}~\bibnamefont
  {Hatzopoulos}}\ and\ \bibinfo {author} {\bibfnamefont {H.~J.}\ \bibnamefont
  {Jensen}},\ }\href@noop {} {\bibfield  {journal} {\bibinfo  {journal}
  {Physical Review E}\ }\textbf {\bibinfo {volume} {78}},\ \bibinfo {pages}
  {011904} (\bibinfo {year} {2008})}\BibitemShut {NoStop}%
\bibitem [{\citenamefont {Pusch}\ \emph {et~al.}(2008)\citenamefont {Pusch},
  \citenamefont {Weber},\ and\ \citenamefont {Porto}}]{pusch2008impact}%
  \BibitemOpen
  \bibfield  {author} {\bibinfo {author} {\bibfnamefont {A.}~\bibnamefont
  {Pusch}}, \bibinfo {author} {\bibfnamefont {S.}~\bibnamefont {Weber}}, \ and\
  \bibinfo {author} {\bibfnamefont {M.}~\bibnamefont {Porto}},\ }\href@noop {}
  {\bibfield  {journal} {\bibinfo  {journal} {Physical Review E}\ }\textbf
  {\bibinfo {volume} {77}},\ \bibinfo {pages} {036120} (\bibinfo {year}
  {2008})}\BibitemShut {NoStop}%
\bibitem [{\citenamefont {Chen}\ and\ \citenamefont
  {Wang}(2008)}]{chen2008promotion}%
  \BibitemOpen
  \bibfield  {author} {\bibinfo {author} {\bibfnamefont {X.}~\bibnamefont
  {Chen}}\ and\ \bibinfo {author} {\bibfnamefont {L.}~\bibnamefont {Wang}},\
  }\href@noop {} {\bibfield  {journal} {\bibinfo  {journal} {Physical Review
  E}\ }\textbf {\bibinfo {volume} {77}},\ \bibinfo {pages} {017103} (\bibinfo
  {year} {2008})}\BibitemShut {NoStop}%
\bibitem [{\citenamefont {Assenza}\ \emph {et~al.}(2008)\citenamefont
  {Assenza}, \citenamefont {G{\'o}mez-Garde{\~n}es},\ and\ \citenamefont
  {Latora}}]{assenza2008enhancement}%
  \BibitemOpen
  \bibfield  {author} {\bibinfo {author} {\bibfnamefont {S.}~\bibnamefont
  {Assenza}}, \bibinfo {author} {\bibfnamefont {J.}~\bibnamefont
  {G{\'o}mez-Garde{\~n}es}}, \ and\ \bibinfo {author} {\bibfnamefont
  {V.}~\bibnamefont {Latora}},\ }\href@noop {} {\bibfield  {journal} {\bibinfo
  {journal} {Physical Review E}\ }\textbf {\bibinfo {volume} {78}},\ \bibinfo
  {pages} {017101} (\bibinfo {year} {2008})}\BibitemShut {NoStop}%
\bibitem [{\citenamefont {Yang}\ \emph {et~al.}(2009)\citenamefont {Yang},
  \citenamefont {Wang}, \citenamefont {Wu}, \citenamefont {Lai},\ and\
  \citenamefont {Wang}}]{yang2009diversity}%
  \BibitemOpen
  \bibfield  {author} {\bibinfo {author} {\bibfnamefont {H.-X.}\ \bibnamefont
  {Yang}}, \bibinfo {author} {\bibfnamefont {W.-X.}\ \bibnamefont {Wang}},
  \bibinfo {author} {\bibfnamefont {Z.-X.}\ \bibnamefont {Wu}}, \bibinfo
  {author} {\bibfnamefont {Y.-C.}\ \bibnamefont {Lai}}, \ and\ \bibinfo
  {author} {\bibfnamefont {B.-H.}\ \bibnamefont {Wang}},\ }\href@noop {}
  {\bibfield  {journal} {\bibinfo  {journal} {Physical Review E}\ }\textbf
  {\bibinfo {volume} {79}},\ \bibinfo {pages} {056107} (\bibinfo {year}
  {2009})}\BibitemShut {NoStop}%
\bibitem [{\citenamefont {G{\'o}mez-Gardenes}\ \emph
  {et~al.}(2012{\natexlab{a}})\citenamefont {G{\'o}mez-Gardenes}, \citenamefont
  {Reinares}, \citenamefont {Arenas},\ and\ \citenamefont
  {Flor{\'\i}a}}]{gomez2012evolution}%
  \BibitemOpen
  \bibfield  {author} {\bibinfo {author} {\bibfnamefont {J.}~\bibnamefont
  {G{\'o}mez-Gardenes}}, \bibinfo {author} {\bibfnamefont {I.}~\bibnamefont
  {Reinares}}, \bibinfo {author} {\bibfnamefont {A.}~\bibnamefont {Arenas}}, \
  and\ \bibinfo {author} {\bibfnamefont {L.~M.}\ \bibnamefont {Flor{\'\i}a}},\
  }\href@noop {} {\bibfield  {journal} {\bibinfo  {journal} {Scientific
  {R}eports - {N}ature}\ }\textbf {\bibinfo {volume} {2}},\ \bibinfo {pages}
  {620} (\bibinfo {year} {2012}{\natexlab{a}})}\BibitemShut {NoStop}%
\bibitem [{\citenamefont {Wang}\ \emph
  {et~al.}(2012{\natexlab{a}})\citenamefont {Wang}, \citenamefont {Chen},\ and\
  \citenamefont {Wang}}]{wang2012probabilistic}%
  \BibitemOpen
  \bibfield  {author} {\bibinfo {author} {\bibfnamefont {B.}~\bibnamefont
  {Wang}}, \bibinfo {author} {\bibfnamefont {X.}~\bibnamefont {Chen}}, \ and\
  \bibinfo {author} {\bibfnamefont {L.}~\bibnamefont {Wang}},\ }\href@noop {}
  {\bibfield  {journal} {\bibinfo  {journal} {Journal of {S}tatistical
  {M}echanics: {T}heory and {E}xperiment}\ }\textbf {\bibinfo {volume}
  {2012}},\ \bibinfo {pages} {P11017} (\bibinfo {year}
  {2012}{\natexlab{a}})}\BibitemShut {NoStop}%
\bibitem [{\citenamefont {G{\'o}mez-Gardenes}\ \emph
  {et~al.}(2012{\natexlab{b}})\citenamefont {G{\'o}mez-Gardenes}, \citenamefont
  {Gracia-L{\'a}zaro}, \citenamefont {Flor{\'\i}a},\ and\ \citenamefont
  {Moreno}}]{gomez2012evolutionary}%
  \BibitemOpen
  \bibfield  {author} {\bibinfo {author} {\bibfnamefont {J.}~\bibnamefont
  {G{\'o}mez-Gardenes}}, \bibinfo {author} {\bibfnamefont {C.}~\bibnamefont
  {Gracia-L{\'a}zaro}}, \bibinfo {author} {\bibfnamefont {L.~M.}\ \bibnamefont
  {Flor{\'\i}a}}, \ and\ \bibinfo {author} {\bibfnamefont {Y.}~\bibnamefont
  {Moreno}},\ }\href@noop {} {\bibfield  {journal} {\bibinfo  {journal}
  {Physical {R}eview {E}}\ }\textbf {\bibinfo {volume} {86}},\ \bibinfo {pages}
  {056113} (\bibinfo {year} {2012}{\natexlab{b}})}\BibitemShut {NoStop}%
\bibitem [{\citenamefont {Wang}\ \emph
  {et~al.}(2012{\natexlab{b}})\citenamefont {Wang}, \citenamefont {Szolnoki},\
  and\ \citenamefont {Perc}}]{wang2012evolution}%
  \BibitemOpen
  \bibfield  {author} {\bibinfo {author} {\bibfnamefont {Z.}~\bibnamefont
  {Wang}}, \bibinfo {author} {\bibfnamefont {A.}~\bibnamefont {Szolnoki}}, \
  and\ \bibinfo {author} {\bibfnamefont {M.}~\bibnamefont {Perc}},\ }\href@noop
  {} {\bibfield  {journal} {\bibinfo  {journal} {{E}urophysics {L}etters)}\
  }\textbf {\bibinfo {volume} {97}},\ \bibinfo {pages} {48001} (\bibinfo {year}
  {2012}{\natexlab{b}})}\BibitemShut {NoStop}%
\bibitem [{\citenamefont {Wang}\ \emph
  {et~al.}(2013{\natexlab{a}})\citenamefont {Wang}, \citenamefont {Szolnoki},\
  and\ \citenamefont {Perc}}]{wang2013optimal}%
  \BibitemOpen
  \bibfield  {author} {\bibinfo {author} {\bibfnamefont {Z.}~\bibnamefont
  {Wang}}, \bibinfo {author} {\bibfnamefont {A.}~\bibnamefont {Szolnoki}}, \
  and\ \bibinfo {author} {\bibfnamefont {M.}~\bibnamefont {Perc}},\ }\href@noop
  {} {\bibfield  {journal} {\bibinfo  {journal} {Scientific {R}eports -
  {N}ature}\ }\textbf {\bibinfo {volume} {3}},\ \bibinfo {pages} {2470}
  (\bibinfo {year} {2013}{\natexlab{a}})}\BibitemShut {NoStop}%
\bibitem [{\citenamefont {Wang}\ \emph
  {et~al.}(2013{\natexlab{b}})\citenamefont {Wang}, \citenamefont {Szolnoki},\
  and\ \citenamefont {Perc}}]{wang2013interdependent}%
  \BibitemOpen
  \bibfield  {author} {\bibinfo {author} {\bibfnamefont {Z.}~\bibnamefont
  {Wang}}, \bibinfo {author} {\bibfnamefont {A.}~\bibnamefont {Szolnoki}}, \
  and\ \bibinfo {author} {\bibfnamefont {M.}~\bibnamefont {Perc}},\ }\href@noop
  {} {\bibfield  {journal} {\bibinfo  {journal} {Scientific {R}eports -
  {N}ature}\ }\textbf {\bibinfo {volume} {3}},\ \bibinfo {pages} {1183}
  (\bibinfo {year} {2013}{\natexlab{b}})}\BibitemShut {NoStop}%
\bibitem [{\citenamefont {Jiang}\ and\ \citenamefont
  {Perc}(2013)}]{jiang2013spreading}%
  \BibitemOpen
  \bibfield  {author} {\bibinfo {author} {\bibfnamefont {L.-L.}\ \bibnamefont
  {Jiang}}\ and\ \bibinfo {author} {\bibfnamefont {M.}~\bibnamefont {Perc}},\
  }\href@noop {} {\bibfield  {journal} {\bibinfo  {journal} {Scientific
  {R}eports - {N}ature}\ ,\ \bibinfo {pages} {2483}} (\bibinfo {year}
  {2013})}\BibitemShut {NoStop}%
\bibitem [{\citenamefont {Santos}\ \emph {et~al.}(2014)\citenamefont {Santos},
  \citenamefont {Dorogovtsev},\ and\ \citenamefont
  {Mendes}}]{santos2014biased}%
  \BibitemOpen
  \bibfield  {author} {\bibinfo {author} {\bibfnamefont {M.~D.}\ \bibnamefont
  {Santos}}, \bibinfo {author} {\bibfnamefont {S.~N.}\ \bibnamefont
  {Dorogovtsev}}, \ and\ \bibinfo {author} {\bibfnamefont {J.~F.~F.}\
  \bibnamefont {Mendes}},\ }\href@noop {} {\bibfield  {journal} {\bibinfo
  {journal} {Scientific {R}eports - {N}ature}\ ,\ \bibinfo {pages} {4436}}
  (\bibinfo {year} {2014})}\BibitemShut {NoStop}%
\bibitem [{\citenamefont {Wang}\ \emph {et~al.}(2014)\citenamefont {Wang},
  \citenamefont {Wang},\ and\ \citenamefont {Perc}}]{wang2014degree}%
  \BibitemOpen
  \bibfield  {author} {\bibinfo {author} {\bibfnamefont {Z.}~\bibnamefont
  {Wang}}, \bibinfo {author} {\bibfnamefont {L.}~\bibnamefont {Wang}}, \ and\
  \bibinfo {author} {\bibfnamefont {M.}~\bibnamefont {Perc}},\ }\href@noop {}
  {\bibfield  {journal} {\bibinfo  {journal} {Physical {R}eview {E}}\ }\textbf
  {\bibinfo {volume} {89}},\ \bibinfo {pages} {052813} (\bibinfo {year}
  {2014})}\BibitemShut {NoStop}%
\bibitem [{\citenamefont {Jin}\ \emph {et~al.}(2014)\citenamefont {Jin},
  \citenamefont {Wang}, \citenamefont {Xia},\ and\ \citenamefont
  {Wang}}]{jin2014spontaneous}%
  \BibitemOpen
  \bibfield  {author} {\bibinfo {author} {\bibfnamefont {Q.}~\bibnamefont
  {Jin}}, \bibinfo {author} {\bibfnamefont {L.}~\bibnamefont {Wang}}, \bibinfo
  {author} {\bibfnamefont {C.-Y.}\ \bibnamefont {Xia}}, \ and\ \bibinfo
  {author} {\bibfnamefont {Z.}~\bibnamefont {Wang}},\ }\href@noop {} {\bibfield
   {journal} {\bibinfo  {journal} {Scientific {R}eports - {N}ature}\ }\textbf
  {\bibinfo {volume} {4}},\ \bibinfo {pages} {4095} (\bibinfo {year}
  {2014})}\BibitemShut {NoStop}%
\bibitem [{\citenamefont {Fu}\ \emph {et~al.}(2008)\citenamefont {Fu},
  \citenamefont {Hauert}, \citenamefont {Nowak},\ and\ \citenamefont
  {L.~Wang}}]{fu2008reputation}%
  \BibitemOpen
  \bibfield  {author} {\bibinfo {author} {\bibfnamefont {F.}~\bibnamefont
  {Fu}}, \bibinfo {author} {\bibfnamefont {C.}~\bibnamefont {Hauert}}, \bibinfo
  {author} {\bibfnamefont {M.}~\bibnamefont {Nowak}}, \ and\ \bibinfo {author}
  {\bibfnamefont {L.}~\bibnamefont {L.~Wang}},\ }\href@noop {} {\bibfield
  {journal} {\bibinfo  {journal} {{P}hysical {R}eview {E}}\ }\textbf {\bibinfo
  {volume} {78}},\ \bibinfo {pages} {026117} (\bibinfo {year}
  {2008})}\BibitemShut {NoStop}%
\bibitem [{\citenamefont {Szab{\'o}}\ and\ \citenamefont
  {Hauert}(2002)}]{szabo2002evolutionary}%
  \BibitemOpen
  \bibfield  {author} {\bibinfo {author} {\bibfnamefont {G.}~\bibnamefont
  {Szab{\'o}}}\ and\ \bibinfo {author} {\bibfnamefont {C.}~\bibnamefont
  {Hauert}},\ }\href@noop {} {\bibfield  {journal} {\bibinfo  {journal}
  {{P}hysical {R}eview {E}}\ }\textbf {\bibinfo {volume} {66}},\ \bibinfo
  {pages} {062903} (\bibinfo {year} {2002})}\BibitemShut {NoStop}%
\bibitem [{\citenamefont {Hauert}\ \emph {et~al.}(2002)\citenamefont {Hauert},
  \citenamefont {Monte}, \citenamefont {Hofbauer},\ and\ \citenamefont
  {Sigmund}}]{hauert2002volunteering}%
  \BibitemOpen
  \bibfield  {author} {\bibinfo {author} {\bibfnamefont {C.}~\bibnamefont
  {Hauert}}, \bibinfo {author} {\bibfnamefont {S.~D.}\ \bibnamefont {Monte}},
  \bibinfo {author} {\bibfnamefont {J.}~\bibnamefont {Hofbauer}}, \ and\
  \bibinfo {author} {\bibfnamefont {K.}~\bibnamefont {Sigmund}},\ }\href@noop
  {} {\bibfield  {journal} {\bibinfo  {journal} {{S}cience}\ }\textbf {\bibinfo
  {volume} {296}},\ \bibinfo {pages} {1129} (\bibinfo {year}
  {2002})}\BibitemShut {NoStop}%
\bibitem [{\citenamefont {Perc}\ and\ \citenamefont
  {Szolnoki}(2008)}]{perc2008social}%
  \BibitemOpen
  \bibfield  {author} {\bibinfo {author} {\bibfnamefont {M.}~\bibnamefont
  {Perc}}\ and\ \bibinfo {author} {\bibfnamefont {A.}~\bibnamefont
  {Szolnoki}},\ }\href@noop {} {\bibfield  {journal} {\bibinfo  {journal}
  {Physical {R}eview {E}}\ }\textbf {\bibinfo {volume} {77}},\ \bibinfo {pages}
  {011904} (\bibinfo {year} {2008})}\BibitemShut {NoStop}%
\bibitem [{\citenamefont {Wang}\ \emph {et~al.}(2006)\citenamefont {Wang},
  \citenamefont {Ren}, \citenamefont {Chen},\ and\ \citenamefont
  {Wang}}]{wang2006memory}%
  \BibitemOpen
  \bibfield  {author} {\bibinfo {author} {\bibfnamefont {W.-X.}\ \bibnamefont
  {Wang}}, \bibinfo {author} {\bibfnamefont {J.}~\bibnamefont {Ren}}, \bibinfo
  {author} {\bibfnamefont {G.}~\bibnamefont {Chen}}, \ and\ \bibinfo {author}
  {\bibfnamefont {B.-H.}\ \bibnamefont {Wang}},\ }\href@noop {} {\bibfield
  {journal} {\bibinfo  {journal} {{P}hysical {R}eview {E}}\ }\textbf {\bibinfo
  {volume} {74}},\ \bibinfo {pages} {056113} (\bibinfo {year}
  {2006})}\BibitemShut {NoStop}%
\bibitem [{\citenamefont {Meloni}\ \emph {et~al.}(2009)\citenamefont {Meloni},
  \citenamefont {Buscarino}, \citenamefont {Fortuna}, \citenamefont {Frasca},
  \citenamefont {G{\'o}mez-Garde{\~n}es}, \citenamefont {Latora},\ and\
  \citenamefont {Moreno}}]{meloni2009effects}%
  \BibitemOpen
  \bibfield  {author} {\bibinfo {author} {\bibfnamefont {S.}~\bibnamefont
  {Meloni}}, \bibinfo {author} {\bibfnamefont {A.}~\bibnamefont {Buscarino}},
  \bibinfo {author} {\bibfnamefont {L.}~\bibnamefont {Fortuna}}, \bibinfo
  {author} {\bibfnamefont {M.}~\bibnamefont {Frasca}}, \bibinfo {author}
  {\bibfnamefont {J.}~\bibnamefont {G{\'o}mez-Garde{\~n}es}}, \bibinfo {author}
  {\bibfnamefont {V.}~\bibnamefont {Latora}}, \ and\ \bibinfo {author}
  {\bibfnamefont {Y.}~\bibnamefont {Moreno}},\ }\href@noop {} {\bibfield
  {journal} {\bibinfo  {journal} {{P}hysical {R}eview {E}}\ }\textbf {\bibinfo
  {volume} {79}},\ \bibinfo {pages} {067101} (\bibinfo {year}
  {2009})}\BibitemShut {NoStop}%
\bibitem [{\citenamefont {Wu}\ and\ \citenamefont
  {Holme}(2009)}]{wu2009effects}%
  \BibitemOpen
  \bibfield  {author} {\bibinfo {author} {\bibfnamefont {Z.-X.}\ \bibnamefont
  {Wu}}\ and\ \bibinfo {author} {\bibfnamefont {P.}~\bibnamefont {Holme}},\
  }\href@noop {} {\bibfield  {journal} {\bibinfo  {journal} {{P}hysical
  {R}eview {E}}\ }\textbf {\bibinfo {volume} {80}},\ \bibinfo {pages} {026108}
  (\bibinfo {year} {2009})}\BibitemShut {NoStop}%
\bibitem [{\citenamefont {Jiang}\ \emph {et~al.}(2010)\citenamefont {Jiang},
  \citenamefont {Wang}, \citenamefont {Lai},\ and\ \citenamefont
  {Wang}}]{jiang2010role}%
  \BibitemOpen
  \bibfield  {author} {\bibinfo {author} {\bibfnamefont {L.-L.}\ \bibnamefont
  {Jiang}}, \bibinfo {author} {\bibfnamefont {W.-X.}\ \bibnamefont {Wang}},
  \bibinfo {author} {\bibfnamefont {Y.-C.}\ \bibnamefont {Lai}}, \ and\
  \bibinfo {author} {\bibfnamefont {B.-H.}\ \bibnamefont {Wang}},\ }\href@noop
  {} {\bibfield  {journal} {\bibinfo  {journal} {Physical Review E}\ }\textbf
  {\bibinfo {volume} {81}},\ \bibinfo {pages} {036108} (\bibinfo {year}
  {2010})}\BibitemShut {NoStop}%
\bibitem [{\citenamefont {McAvoy}\ and\ \citenamefont
  {Hauert}(2015)}]{mcavoy2015asymmetric}%
  \BibitemOpen
  \bibfield  {author} {\bibinfo {author} {\bibfnamefont {A.}~\bibnamefont
  {McAvoy}}\ and\ \bibinfo {author} {\bibfnamefont {C.}~\bibnamefont
  {Hauert}},\ }\href@noop {} {\bibfield  {journal} {\bibinfo  {journal} {PLoS
  computational biology}\ }\textbf {\bibinfo {volume} {11}},\ \bibinfo {pages}
  {e1004349} (\bibinfo {year} {2015})}\BibitemShut {NoStop}%
\bibitem [{\citenamefont {Charlesworth}(1994)}]{charlesworth1994evolution}%
  \BibitemOpen
  \bibfield  {author} {\bibinfo {author} {\bibfnamefont {B.}~\bibnamefont
  {Charlesworth}},\ }\href@noop {} {\emph {\bibinfo {title} {Evolution in
  age-structured populations}}},\ Vol.~\bibinfo {volume} {2}\ (\bibinfo
  {publisher} {Cambridge University Press Cambridge},\ \bibinfo {year}
  {1994})\BibitemShut {NoStop}%
\bibitem [{\citenamefont {Perc}\ and\ \citenamefont
  {Szolnoki}(2010)}]{perc2010coevolutionary}%
  \BibitemOpen
  \bibfield  {author} {\bibinfo {author} {\bibfnamefont {M.}~\bibnamefont
  {Perc}}\ and\ \bibinfo {author} {\bibfnamefont {A.}~\bibnamefont
  {Szolnoki}},\ }\href@noop {} {\bibfield  {journal} {\bibinfo  {journal}
  {BioSystems}\ }\textbf {\bibinfo {volume} {99}},\ \bibinfo {pages} {109 }
  (\bibinfo {year} {2010})}\BibitemShut {NoStop}%
\bibitem [{\citenamefont {Szolnoki}\ \emph {et~al.}(2009)\citenamefont
  {Szolnoki}, \citenamefont {Perc}, \citenamefont {Szab{\'o}},\ and\
  \citenamefont {Stark}}]{szolnoki2009impact}%
  \BibitemOpen
  \bibfield  {author} {\bibinfo {author} {\bibfnamefont {A.}~\bibnamefont
  {Szolnoki}}, \bibinfo {author} {\bibfnamefont {M.}~\bibnamefont {Perc}},
  \bibinfo {author} {\bibfnamefont {G.}~\bibnamefont {Szab{\'o}}}, \ and\
  \bibinfo {author} {\bibfnamefont {H.-U.}\ \bibnamefont {Stark}},\ }\href@noop
  {} {\bibfield  {journal} {\bibinfo  {journal} {Physical {R}eview {E}}\
  }\textbf {\bibinfo {volume} {80}},\ \bibinfo {pages} {021901} (\bibinfo
  {year} {2009})}\BibitemShut {NoStop}%
\bibitem [{\citenamefont {Wang}\ \emph
  {et~al.}(2012{\natexlab{c}})\citenamefont {Wang}, \citenamefont {Zhu},\ and\
  \citenamefont {Arenzon}}]{wang2012cooperation}%
  \BibitemOpen
  \bibfield  {author} {\bibinfo {author} {\bibfnamefont {Z.}~\bibnamefont
  {Wang}}, \bibinfo {author} {\bibfnamefont {X.}~\bibnamefont {Zhu}}, \ and\
  \bibinfo {author} {\bibfnamefont {J.~J.}\ \bibnamefont {Arenzon}},\
  }\href@noop {} {\bibfield  {journal} {\bibinfo  {journal} {Physical {R}eview
  {E}}\ }\textbf {\bibinfo {volume} {85}},\ \bibinfo {pages} {011149} (\bibinfo
  {year} {2012}{\natexlab{c}})}\BibitemShut {NoStop}%
\bibitem [{\citenamefont {Z.}\ \emph {et~al.}(2012)\citenamefont {Z.},
  \citenamefont {Yang}, \citenamefont {Yu},\ and\ \citenamefont
  {Liao}}]{wang2012age}%
  \BibitemOpen
  \bibfield  {author} {\bibinfo {author} {\bibfnamefont {Z.~W.}\ \bibnamefont
  {Z.}}, \bibinfo {author} {\bibfnamefont {Y.-H.}\ \bibnamefont {Yang}},
  \bibinfo {author} {\bibfnamefont {M.-X.}\ \bibnamefont {Yu}}, \ and\ \bibinfo
  {author} {\bibfnamefont {L.-G.}\ \bibnamefont {Liao}},\ }\href@noop {}
  {\bibfield  {journal} {\bibinfo  {journal} {International Journal of Modern
  Physics C}\ }\textbf {\bibinfo {volume} {23}},\ \bibinfo {pages} {1250013}
  (\bibinfo {year} {2012})}\BibitemShut {NoStop}%
\bibitem [{\citenamefont {Felsenthal}(2018)}]{time2018}%
  \BibitemOpen
  \bibfield  {author} {\bibinfo {author} {\bibfnamefont {E.}~\bibnamefont
  {Felsenthal}},\ }\href@noop {} {\bibfield  {journal} {\bibinfo  {journal}
  {Time Magazine}\ }\textbf {\bibinfo {volume} {April 30}} (\bibinfo {year}
  {2018})}\BibitemShut {NoStop}%
\bibitem [{\citenamefont {Spisak}(2012)}]{spisak2012general}%
  \BibitemOpen
  \bibfield  {author} {\bibinfo {author} {\bibfnamefont {B.~R.}\ \bibnamefont
  {Spisak}},\ }\href@noop {} {\bibfield  {journal} {\bibinfo  {journal} {PLoS
  One}\ }\textbf {\bibinfo {volume} {7}},\ \bibinfo {pages} {e36945} (\bibinfo
  {year} {2012})}\BibitemShut {NoStop}%
\bibitem [{\citenamefont {Jacobs}\ \emph {et~al.}(2008)\citenamefont {Jacobs},
  \citenamefont {Maumy},\ and\ \citenamefont {Petit}}]{jacobs2008influence}%
  \BibitemOpen
  \bibfield  {author} {\bibinfo {author} {\bibfnamefont {A.}~\bibnamefont
  {Jacobs}}, \bibinfo {author} {\bibfnamefont {M.}~\bibnamefont {Maumy}}, \
  and\ \bibinfo {author} {\bibfnamefont {O.}~\bibnamefont {Petit}},\
  }\href@noop {} {\bibfield  {journal} {\bibinfo  {journal} {Behavioural
  Processes}\ }\textbf {\bibinfo {volume} {79}},\ \bibinfo {pages} {111}
  (\bibinfo {year} {2008})}\BibitemShut {NoStop}%
\bibitem [{\citenamefont {Watanabe}(2008)}]{watanabe2008review}%
  \BibitemOpen
  \bibfield  {author} {\bibinfo {author} {\bibfnamefont {K.}~\bibnamefont
  {Watanabe}},\ }in\ \href@noop {} {\emph {\bibinfo {booktitle} {Primate
  origins of human cognition and behavior}}}\ (\bibinfo  {publisher}
  {Springer},\ \bibinfo {year} {2008})\ pp.\ \bibinfo {pages}
  {405--417}\BibitemShut {NoStop}%
\bibitem [{\citenamefont {Penna}(1995)}]{penna1995bit}%
  \BibitemOpen
  \bibfield  {author} {\bibinfo {author} {\bibfnamefont {T.~J.}\ \bibnamefont
  {Penna}},\ }\href@noop {} {\bibfield  {journal} {\bibinfo  {journal} {Journal
  of Statistical Physics}\ }\textbf {\bibinfo {volume} {78}},\ \bibinfo {pages}
  {1629} (\bibinfo {year} {1995})}\BibitemShut {NoStop}%
\bibitem [{\citenamefont {Chen}\ \emph {et~al.}(2003)\citenamefont {Chen},
  \citenamefont {Chen}, \citenamefont {Sun},\ and\ \citenamefont
  {Chen}}]{chen2003life}%
  \BibitemOpen
  \bibfield  {author} {\bibinfo {author} {\bibfnamefont {C.~C.}\ \bibnamefont
  {Chen}}, \bibinfo {author} {\bibfnamefont {Y.-T.}\ \bibnamefont {Chen}},
  \bibinfo {author} {\bibfnamefont {Y.}~\bibnamefont {Sun}}, \ and\ \bibinfo
  {author} {\bibfnamefont {M.~C.}\ \bibnamefont {Chen}},\ }in\ \href@noop {}
  {\emph {\bibinfo {booktitle} {European Conference on Machine Learning}}}\
  (\bibinfo {organization} {Springer},\ \bibinfo {year} {2003})\ pp.\ \bibinfo
  {pages} {47--59}\BibitemShut {NoStop}%
\bibitem [{\citenamefont {Avery}\ \emph {et~al.}(2007)\citenamefont {Avery},
  \citenamefont {McKay},\ and\ \citenamefont {Wilson}}]{avery2007engaging}%
  \BibitemOpen
  \bibfield  {author} {\bibinfo {author} {\bibfnamefont {D.~R.}\ \bibnamefont
  {Avery}}, \bibinfo {author} {\bibfnamefont {P.~F.}\ \bibnamefont {McKay}}, \
  and\ \bibinfo {author} {\bibfnamefont {D.~C.}\ \bibnamefont {Wilson}},\
  }\href@noop {} {\bibfield  {journal} {\bibinfo  {journal} {Journal of Applied
  Psychology}\ }\textbf {\bibinfo {volume} {92}},\ \bibinfo {pages} {1542}
  (\bibinfo {year} {2007})}\BibitemShut {NoStop}%
\bibitem [{\citenamefont {Wiersema}\ and\ \citenamefont
  {Bantel}(1992)}]{wiersema1992top}%
  \BibitemOpen
  \bibfield  {author} {\bibinfo {author} {\bibfnamefont {M.~F.}\ \bibnamefont
  {Wiersema}}\ and\ \bibinfo {author} {\bibfnamefont {K.~A.}\ \bibnamefont
  {Bantel}},\ }\href@noop {} {\bibfield  {journal} {\bibinfo  {journal}
  {Academy of Management Journal}\ }\textbf {\bibinfo {volume} {35}},\ \bibinfo
  {pages} {91} (\bibinfo {year} {1992})}\BibitemShut {NoStop}%
\bibitem [{\citenamefont {der Vegt}(2002)}]{van2002effects}%
  \BibitemOpen
  \bibfield  {author} {\bibinfo {author} {\bibfnamefont {G.~S.~V.}\
  \bibnamefont {der Vegt}},\ }\href@noop {} {\bibfield  {journal} {\bibinfo
  {journal} {Journal of Occupational and Organizational Psychology}\ }\textbf
  {\bibinfo {volume} {75}},\ \bibinfo {pages} {439} (\bibinfo {year}
  {2002})}\BibitemShut {NoStop}%
\bibitem [{\citenamefont {Maccoby}(2004)}]{maccoby2004people}%
  \BibitemOpen
  \bibfield  {author} {\bibinfo {author} {\bibfnamefont {M.}~\bibnamefont
  {Maccoby}},\ }\href@noop {} {\bibfield  {journal} {\bibinfo  {journal}
  {Harvard Business Review}\ }\textbf {\bibinfo {volume} {82}},\ \bibinfo
  {pages} {76} (\bibinfo {year} {2004})}\BibitemShut {NoStop}%
\bibitem [{\citenamefont {Braha}\ and\ \citenamefont
  {de~Aguiar}(2017)}]{braha2017voting}%
  \BibitemOpen
  \bibfield  {author} {\bibinfo {author} {\bibfnamefont {D.}~\bibnamefont
  {Braha}}\ and\ \bibinfo {author} {\bibfnamefont {M.~A.~M.}\ \bibnamefont
  {de~Aguiar}},\ }\href@noop {} {\bibfield  {journal} {\bibinfo  {journal}
  {PloS One}\ }\textbf {\bibinfo {volume} {12}},\ \bibinfo {pages} {e0177970}
  (\bibinfo {year} {2017})}\BibitemShut {NoStop}%
\bibitem [{\citenamefont {Szab{\'o}}\ and\ \citenamefont
  {T{\H{o}}ke}(1998)}]{szabo1998evolutionary}%
  \BibitemOpen
  \bibfield  {author} {\bibinfo {author} {\bibfnamefont {G.}~\bibnamefont
  {Szab{\'o}}}\ and\ \bibinfo {author} {\bibfnamefont {C.}~\bibnamefont
  {T{\H{o}}ke}},\ }\href@noop {} {\bibfield  {journal} {\bibinfo  {journal}
  {Physical {R}eview {E}}\ }\textbf {\bibinfo {volume} {58}},\ \bibinfo {pages}
  {69} (\bibinfo {year} {1998})}\BibitemShut {NoStop}%
\bibitem [{\citenamefont {Harris}(1974)}]{harris}%
  \BibitemOpen
  \bibfield  {author} {\bibinfo {author} {\bibfnamefont {A.~B.}\ \bibnamefont
  {Harris}},\ }\href@noop {} {\bibfield  {journal} {\bibinfo  {journal}
  {Journal of Physics C: Solid State Physics}\ }\textbf {\bibinfo {volume}
  {7}},\ \bibinfo {pages} {1671} (\bibinfo {year} {1974})}\BibitemShut
  {NoStop}%
\bibitem [{\citenamefont {Moreira}\ and\ \citenamefont
  {Dickman}(1996)}]{dickman}%
  \BibitemOpen
  \bibfield  {author} {\bibinfo {author} {\bibfnamefont {A.~G.}\ \bibnamefont
  {Moreira}}\ and\ \bibinfo {author} {\bibfnamefont {R.}~\bibnamefont
  {Dickman}},\ }\href@noop {} {\bibfield  {journal} {\bibinfo  {journal}
  {Physical Review E}\ }\textbf {\bibinfo {volume} {54}},\ \bibinfo {pages}
  {R3090} (\bibinfo {year} {1996})}\BibitemShut {NoStop}%
\bibitem [{\citenamefont {de~Andrade}\ and\ \citenamefont
  {Figueiredo}(2016)}]{wagner}%
  \BibitemOpen
  \bibfield  {author} {\bibinfo {author} {\bibfnamefont {M.}~\bibnamefont
  {de~Andrade}}\ and\ \bibinfo {author} {\bibfnamefont {W.}~\bibnamefont
  {Figueiredo}},\ }\href@noop {} {\bibfield  {journal} {\bibinfo  {journal}
  {Physics Letters A}\ }\textbf {\bibinfo {volume} {380}},\ \bibinfo {pages}
  {2628} (\bibinfo {year} {2016})}\BibitemShut {NoStop}%
\end{thebibliography}%
\end{document}